# Une approche modulaire probabiliste pour le routage à QdS intégrée.

**Said Hoceini, Abdelhamid Mellouk, Hayet Hafi**

*LiSSi, IUT Créteil-Vitry, Université Paris XII*
122 rue Paul Armangot Vitry sur Seine 94400 France
mellouk@univ-paris12.fr

*RESUME. Les techniques traditionnelles de routage sont de plus en plus mal adapteés aux réseaux actuels. En effet, leur manque de réactivité vis à vis de la nature du trafic et au caractère variable des caractéristiques du réseau de transport les rendent souvent difficilement exploitables ou alors au prix d'un surdimensionnement des ressources du réseau (bande passante, mémoire tampon, charge CPU, etc.). L'objectif de cet article est de proposer un nouveau modèle algorithmique permettant de traiter la problématique du routage adaptatif dans un réseau de communication à trafic irrégulier et fortement dynamique. Celui-ci intègre des critères de QdS aussi bien dynamiques que statiques. Un exemple d'utilisation d'un tel modèle est donné ici : il repose sur l'apprentissage en continu des paramètres de routage, en particulier le temps bout-en-bout de remise des paquets l'état des files d'attente des routeurs. Les performances obtenues, comparativement aux approches classiques, sont très intéressantes dans le cas d'un trafic fortement dynamique.*

*ABSTRACT. Due to emerging real-time and multimedia applications, efficient routing of information packets in dynamically changing communication network requires that as the load levels, traffic patterns and topology of the network change, the routing policy also adapts. We focused in this paper on QoS based routing by developing a neuro-dynamic programming to construct dynamic state dependent routing policies. We propose an approach based on adaptive algorithm for packet routing using reinforcement learning which optimizes two criteria: cumulative cost path and end-to-end delay. Numerical results obtained with OPNET simulator for different packet interarrival times statistical distributions with different levels of traffic's load show that the proposed approach gives better results compared to standard optimal path routing algorithms.*

*MOTS-CLES. Routage multi chemins, Minimisation du délai de bout en bout, Routage avec QoS, Q-Routing, Apprentissage par renforcement.*

**1. Introduction**

Le trafic dans les réseaux actuels se caractérise de plus en plus par des anomalies ayant pour conséquence des changements d'états imprévisibles dans le réseau [2] tout en nécessitant un transport garanti en terme de QdS. Il s'avère donc important de pouvoir s'adapter à la nature variable des paramètres d'un réseau ainsi qu'à la dynamicité des ressources disponibles. C'est ainsi que les approches développées ces dernières années ne se contentent pas seulement de manipuler des informations sur les plans de données, de contrôle et de gestion, elles intègrent de plus en plus des connaissances, acquises ou apprises, sur différents paramètres définissant l'état du réseau : trafic, ressources, besoins, etc.

L'objectif de cet article est de proposer un modèle de routage adaptatif capable de s'adapter aux conditions et à la nature d'un trafic dynamique d'une part et à l'utilisation des ressources du réseau d'autre part.

**2. Le modèle proposé**

Le problème posé ici est la mise en œuvre d'un algorithme de routage multi-chemins. Soit le graphe $G = (X, U)$ consistant en un ensemble $X$ avec $|X| = N$ nœuds et un ensemble $U$ avec $|U| = M$ liens. Les nœuds représentent les routeurs, commutateurs ou relais d'un réseau tandis que les liens représentent les tuyaux de communication (fibre optique, sans fil, etc.). Un lien spécifique de l'ensemble $U$ entre les nœuds $u$ et $v$ est notée $(u, v)$. Chaque lien $(u, v) \in U$ est caratérisé par un vecteur $W$ de paramètres de dimension $m$, tel que $W(u,v)=[w_1(u,v), w_2(u,v),....., w_m(u,v)]$, où $w_m(u,v) \geq 0 \;\; \forall (u,v) \in U$ et les $m$ composants se refèrent aux critèes de la QdS comme le délai, le coût, etc. Un chemin dans $G$ est noté $P(s, t)$ s'il relie le nœud source $s$ au nœud destinataire $d$. Les algorithmes de routage à QdS sont ceux qui calculent le chemin $P$, parmi les $k$ chemins recherchés, optimisant une ou plusieurs contraintes liées à la QdS. Le vecteur $L$ sera appelé vecteur de contraintes et les valeurs $L_i$ mesurent les paramètres de QdS fixés par l'utilisateur (ou l'application). Les mesures de QdS sont de trois types : additif (ex. délai), multiplicatif (ex. taux de perte) ou min-max (ex. bande passante minimum nécessaire à un flux). On parle alors de typologie de métriques [3]. Au vu des multiples métriques $w_i$ appelées à être optimisées, le problème du routage devient NP-complet dès qu'il s'agit d'optimiser deux critères (ou types de critères) non corrélés à la fois [4]. Le modèle que nous proposons, de complexité $O(kNlog(kN)+k^2mM)$, agit en deux étapes. Dans un premier temps, un ensemble de chemins potentiels est sélectionné au regard d'une fonction de coût à optimiser construite sur la base de critères statiques comme la bande passante, le nombre de sauts, le délai de bout en bout et le taux d'erreur (1).

$$\text{Coût}_{statique} = f \text{ (bande passante, nombre de sauts, délai, taux d'erreur, etc.)} \quad (1)$$

Ensuite, le flux est distribué sur l'ensemble des chemins potentiels et le taux d'utilisation de chaque chemin est mis à jour en temps réel. Le coût de chaque chemin sélectionné dépend de l'évaluation des critères dynamiques tels que le taux d'acceptation, le taux de perte de paquet, le délai réel, la gigue, etc. Ceci exige également la définition d'une fonction de coût basée sur des critères dynamiques qui reste toujours un problème d'optimisation multi-critères (2).

$Coût_{dynamique}$= f'(disponibilité, taux de perte de paquets, délai mesuré, gigue, bande passante mesurée, etc.) (2)

Nous allons détailler dans la suite une application du modèle proposé. Elle consiste dans le développement d'un algorithme basé sur un mécanisme d'apprentissage, issu du Q Learning, et permettant d'optimiser deux type de critères de QdS : un critère statique (le coût du lien) et un critère dynamique (le délai mesuré)

## 3. Application du modèle pour l'optimisation de deux critères : le coût et le délai de bout-en-bout.

Notre approche algorithmique est basée sur la technique du routage multi-chemins combiné avec le Q-Learning. L'espace d'exploration est réduit aux *K* meilleurs chemins au sens de critères statiques qui peuvent être liés par exemple à la bande passante, au coût des liens, au délai mesuré ou au taux de perte. Dans ce travail, nous nous sommes focalisés sur la recherche des chemins minimisant deux critères : le coût (pour des questions de simplicité, nous considérons des coûts unitaires) et le délai de bout en bout. Pour ce faire, nous utilisons l'algorithme de Dijkstra généralisé [5] afin de trouver les *K* meilleurs chemins sur la base des coûts des liens. La répartition du trafic se fera ensuite sur l'ensemble de ces *K* chemins en fonction du meilleur temps d'acheminement de bout en bout, celui-ci étant calculé par le biais d'un mécanisme basé sur l'apprentissage par renforcement. L'algorithme de mise à jour des paramètres de routage repose sur une méthode hybride associant le principe de l'exploration avancée à chaque fois qu'un paquet de données est échangé entre routeurs, à celui de l'exploration probabiliste permettant d'explorer les *(K -1)* autres chemins sans surcharger le réseau.

Pour résoudre le problème lié à l'exploration du réseau dans la phase d'apprentissage, une solution intéressante consiste à introduire un mécanisme permettant l'exploration régulière de tous les chemins. Pour notre approche, nous avons opté en premier lieu pour une exploration probabiliste (version appelée *KSPQR* de l'algorithme). Elle permet une exploration de temps à autre des (K-1) chemins pré sélectionnés par le premier module. Elle consiste à assigner arbitrairement une probabilité, que nous noterons $P_{max}$, au chemin optimal, les *(K-1)* autres chemins auront une probabilité équivalente à *(1 - $P_{max}$)/K*. Une deuxième technique présentée dans cet article, appelée KOQRA, repose sur un calcul adaptatif de la probabilité de dis-tribution prenant en compte les paramètres qui influent sur le choix du routeur : le délai estimé de bout en bout et le temps d'attente au niveau du routeur (temps de séjour dans la file d'attente). L'idée que nous proposons, issue de la théorie liée à l'intelligence collective des colonies de fourmis, consiste à introduire une procédure adaptative de calcul de probabilités pour chaque chemin, prenant en compte les deux paramètres définis précédemment. En adaptant la valeur de la probabilité, il s'agit de permettre au routeur d'éviter d'envoyer un paquet sur une interface dont le routeur d'extrémité possède une file d'attente saturée, même si le chemin dont ce dernier fait partie, présente le plus court délai.

## 4. Simulations

Les simulations, effectuées sur la plate forme OPNET, ont porté sur l'architecture japonaise *NTTnet* présenté dans [6] (à droite de la figure 1). Pour un but de comparaison, les performances des deux versions de l'algorithme (KSPQR et KOQRA) ont été comparées à ceux des algorithmes traditionnels SPF (*Shortest Path First*) et SOMR (*Standard Optimal Multi-Path Routing*) en termes de temps d'acheminement moyen des paquets.

L'analyse des caractéristiques s'est faite au niveau de l'entité la plus fine qu'est le paquet en utilisant un modèle poissonien pour simuler le trafic. L'ensemble de nos évaluations sont faites en dehors de tout contrôle supplémentaire se faisant au-delà de la couche réseau comme par exemple un contrôle d'erreurs ou un ordonnacement de paquets par la couche transport. Des travaux futurs doivent nécessairement les prendre en compte dans le cas d'un déploiement à grande échelle et d'une étude portant sur leur interaction avec l'ensemble des composants d'un réseau.

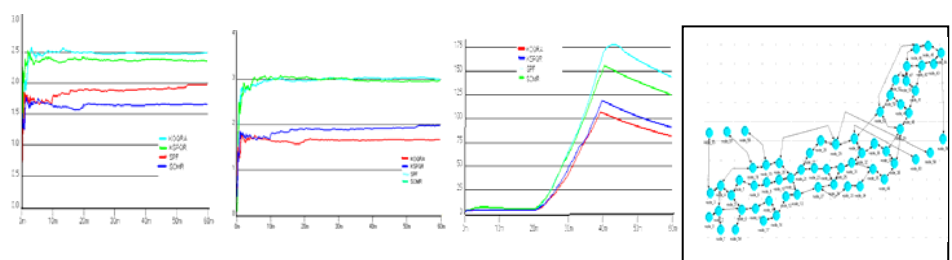

**Figure 1 :** Temps moyen d'acheminement pour un trafic (gauche : trafic faible, milieu : trafic fort, droite : pic de trafic) et le réseau NTT.

Sur l'ensemble des figures représentant les résultats des différentes expériences (Figure 1), l'axe des abcisses représente le temps de simulation tandis qu'une indication sur le délai moyen d'acheminement des paquets est représentée au niveau de l'axe des ordonnées. Les résultats de simulation obtenus sur le réseau *NTT*, dans des conditions de trafic faible, et représentés sur la partie gauche de la figure 1, montrent que les temps moyens d'acheminement sont plus favorables aux algorithmes traditionnels SPF et SOMR comparativement aux autres algorithmes basés sur les techniques adaptatives, où on constate que le temps moyen d'acheminement est augmenté de près de 15%. Comme ces dernières sont basées sur un apprentissage continu par le biais des paquets de contrôle qu'elles génèrent, ces derniers surchargent inutilement le réseau et font chuter par conséquent les performances du routage. Dans le cas où le trafic sur le réseau devient très chargé, les résultats illustrés dans la partie centrale de la figure 1 montrent distinctement que les approches adaptatives donnent de meilleures performances en termes de temps moyen d'acheminement des paquets comparativement aux algorithmes classiques (SOMR et SPF). Comme le facteur lié à la congestion est pris en compte d'une façon adaptative, dynamique et qualitative dans les algorithmes KOQRA et KSPQR, ces derniers établissent de nouvelles routes en fonction de leurs qualités (temps d'attente et temps de transit) à chaque fois qu'une route en cours d'utilisation s'avère encombrée. Les algorithmes classiques (SOMR et SPF) mettent quant à eux plus de temps à réagir à cette congestion au regard de la périodicité utilisée dans les mises à jour de leurs paramètres et à leur estimation des paramètres de routage de manière locale. Le même résultat est constaté dans le dernier scénario testé (courbe de droite de la figure 1) où les conditions d'un pic de trafic sont créées.

**6. Conclusion**

Dans cet article, nous avons présenté une nouvelle approche basée sur la technique du routage multi-chemins combiné avec des algorithmes de type adaptatif. Une analyse comparative des performances de l'approche proposée montre clairement l'efficacité du modèle proposé et son intérêt pour des réseaux à forte charge ou soumis à des changements soudains conduisant à des pics sporadiques de trafic. Comme l'optimisation d'une fonction de coût composée de plusieurs critères non corrélés est un problème NP-complet, nous nous proposons comme suite de ce travail d'étudier, sur la base des algorithmes que nous avons développés, les possibilités d'intégrer à la fois plusieurs paramètres de QoS (bande passante et taux de perte des paquets) dans le signal de renforcement.

**7. Bibliographie**